\documentclass[prd,aps,amsfonts,preprint,floats,tightenlines,floatfix,nofootinbib,showpacs]{revtex4}
\usepackage{exscale}                  
\usepackage[intlimits]{amsmath}       
\usepackage{amsfonts}
\usepackage{amssymb,amscd}

\usepackage{epsfig} 

\newcommand{\bi}{\bigskip}
\newcommand{\no}{\noindent}
\newcommand{\be}{\begin{equation}}
\newcommand{\ee}{\end{equation}}

\newcommand{\bea}{\begin{eqnarray}}
\newcommand{\eea}{\end{eqnarray}}
\newcommand{\hk}{\hspace{0.1cm}}

\newcommand{\rk}{\right)}
\newcommand{\lk}{\left(}

\newcommand{\vA}{\vec{A}}

\begin{document}

\title{Confining Solution of the Dyson-Schwinger Equations in Coulomb Gauge}
\author{D. Epple, H. Reinhardt, W. Schleifenbaum}
\affiliation{Institut f\"ur Theoretische Physik\\
Auf der Morgenstelle 14\\
D-72076 T\"ubingen\\
Germany}

\begin{abstract}
The Dyson-Schwinger equations arising from minimizing the vacuum energy density
in the Hamiltonian approach to Yang-Mills theory in Coulomb gauge are solved
numerically. A new solution is presented which gives rise to a strictly
linearly rising static quark potential and whose existence was previously observed
in the infrared analysis of the Dyson-Schwinger equations. For the new solution
we also present the static quark potential and calculate the running coupling
constant from the ghost-gluon vertex. 
\end{abstract}

\maketitle


\section{Introduction}
\bi

\no
Recently, there has been renewed interest in the Hamiltonian approach to
Yang-Mills theory in Coulomb gauge \cite{Christ:1980ku}.
In this gauge the Yang-Mills Schr\"odinger
equation was solved variationally in the continuum using Gaussian types of wave
functionals \cite{Szczepaniak:2001rg,Feuchter:2004mk,Reinhardt:2004mm}. Minimization of the energy density results in a set of
Dyson-Schwinger equations. These equations have been solved analytically in the
infrared using power law ans\"atze for the corresponding Green functions \cite{Schleifenbaum:2006bq}.
Thereby, two infrared solutions of the Dyson-Schwinger equations have been found which differ in the infrared
exponents. Only one of these two solutions was previously 
 found in the numerical solution of the Dyson-Schwinger equations in the whole momentum regime \cite{Feuchter:2004mk}. We present here
the other one which has the attractive property that it gives rise to a strictly
linear rising static quark potential.
\bi

\no
The organization of the paper is as follows: In the next section we briefly
summarize the basic ingredients of the Hamiltonian approach to Yang-Mills theory in
Coulomb gauge and present the resulting Dyson-Schwinger equations. The numerical
solutions of these equations are presented in section 3. We also calculate from
these solutions the static quark potential and the running coupling constant.
Finally, our conclusions are given in section 4.
\bi

\no
\section{The Dyson-Schwinger equations in the Hamiltonian approach to Yang-Mills
theory in Coulomb gauge}
\bi

\no
In Coulomb gauge, where $\vec{\partial}\cdot \vA = 0$, the Yang-Mills Hamiltonian reads
\cite{Christ:1980ku}
\be
\label{G1}
H  =  \frac{1}{2} \int J^{- 1}  \Pi J  \Pi + \frac{1}{2} \int B^2
 + \frac{g^2}{2} \int J^{- 1} \rho (- \hat{D} \partial)^{- 1} (- \partial^2) 
 (- \hat{D} \partial)^{- 1} J \rho \hk .
\ee
Here, $\Pi^a_i (x) = \delta / i \delta A^a_i (x)$ denotes the momentum operator,
representing the color electric field, $B$ is the color magnetic field and 
$\rho^a (x) = - \hat{A}^{a b}_i (x)
\Pi^b_i (x)$ is the non-Abelian color charge of the gauge field. Furthermore, 
$J (A) = Det (- \hat{D}_i \partial_i)$ is the Faddeev-Popov determinant with
$\hat{D}^{a b}_i = \delta^{a b} \partial_i + \hat{A}^{a b}_i$ being the covariant derivative in the
adjoint representation of the gauge group ($\hat{A}^{a b}_i  = f^{a c b}
 A^c_i$ with the structure
constants $f^{a b c}$). 
The first term
in the Hamiltonian is the Laplacian in the curved space of gauge orbits, 
the second term represents
the potential and the last term arises from the longitudinal momentum part of
the kinetic energy after resolving Gauss' law. We use the following
 ansatz for the vacuum wave
functional \cite{Feuchter:2004mk}
\be
\label{G2}
\Psi [A] = {\cal N} \frac{1}{\sqrt{J [A]}}
 \exp \lk - \frac{1}{2} \int A \omega A \rk
\hk ,
\ee
where 
$\omega ( | x - x' |)$ is 
 a variational
kernel determined by minimizing the energy
\be
\label{G3}
\langle \Psi | H | \Psi \rangle = \int {\cal D}
 A J [A] \Psi^* [A] H \Psi [A] \hk .
\ee
This yields a set of coupled Dyson-Schwinger equations for the ghost
propagator
\be
\label{G4}
G^{a b} (x, x') = \langle \Psi | \langle x, a | (- \hat{D} \partial)^{- 1} | x',
b \rangle | \Psi \rangle = \frac{1}{g} (- \partial^2)^{- 1} d (x, x')  \hk ,
\ee
the gluon propagator 
 \be
 \label{G5}
 D^{a b}_{i j} (x, x') = \langle \Psi | A^a_i (x) A^b_j (x') | \Psi \rangle =
 \frac{1}{2} \delta^{a b} t_{i j} (x) \omega^{- 1} (x, x') \hk , \hk t_{i j} (x)
 = \delta_{i j} - \frac{\partial^x_i \partial^x_j}{\partial^2_x} \hk ,
 \ee
and the Coulomb form factor $f$ defined by  
 \bea
 \label{definition-f}
& & \langle \Psi | ( - \hat{D} \partial)^{- 1} (- \partial^2) (- \hat{D}
\partial)^{- 1} | \Psi \rangle  =  \langle \Psi | ( - \hat{D} \partial)^{- 1} | \Psi \rangle (- \partial^2) f
\langle \Psi | ( - \hat{D} \partial)^{- 1} | \Psi \rangle \hk .
 \eea
A further quantity arising in the evaluation of $\langle \Psi | H | \Psi \rangle$ is
the scalar curvature $\chi$, 
 \begin{align}
 \chi(x,x') &= \frac{1}{2 (N^2_C - 1)} t_{i j}(x) \chi^{a a}_{i j}(x,x') \, ,& \chi^{a b}_{i j} (x, x') &= - \frac{1}{2} \left\langle \Psi \left| \frac{\delta^2 \ln
 J [A]}{\delta A^a_i (x) \delta A^b_j (x')} \right| \Psi \right\rangle \hk ,
 \label{G11} 
 \end{align}
which denotes the ghost loop contribution to the gluon energy. This quantity measures
the ``curvature'' of the space of transversal gauge orbits and vanishes in QED.

Following ref.\ \cite{Feuchter:2004mk}, we will solve here the Dyson-Schwinger
 equations to one-loop order. To this order it is sufficient to calculate the
 Coulomb form factor $f$ perturbatively to ensure the proper ultraviolet asymptotics
 of the corresponding loop integrals (see ref.\ \cite{Feuchter:2004mk} for
 more details). In the infrared, this results in $f(k\to 0)=\mathrm{const.}$ After regularization and renormalization using subtracted equations with some arbitrary renormalization scale $\mu$,
 the remaining Dyson-Schwinger equations for the ghost and gluon propagators
 and the expression for the curvature
 in momentum space read \cite{Feuchter:2004mk}
 \bea
\label{G7}
 d^{- 1} (k) & = & d^{- 1} (\mu) - \Delta I_d (k, \mu)\hk , \\
 \label{G8}
 \omega^2 (k)  & = & \chi^2 (k) + k^2 + 2 \chi (k) \lk \Delta I^{(1)}_\omega (k,
 \mu) - \Delta I^{(1)}_\omega (0, \mu) + c \rk + \Delta I^{(2)}_\omega (k, \mu)
 + c_0 \hk ,\label{eq-omega}\\
\chi (k) & = & \chi (\mu) + \Delta I_\chi (k, \mu)\hk ,
 \eea
 with $c_0=\omega^2(\mu)-\chi^2(\mu)-\mu^2-2\chi(\mu)(-\Delta I_\omega^{(1)}(0)+c)$.
 Here $d^{-1}(\mu)$, $\omega(\mu)$, $\chi(\mu)$, and $c$ are
 renormalization constants.
 Furthermore, the $\Delta I_x$ refer to the
 following differences of loop integrals
 \bea
 \label{G9}
 \Delta I_d (k, \mu) & = & I_d (k) - I_d (\mu) \hk , \hk I_d (k) = \frac{N_c}{2}\int\frac{d^3q}{(2\pi)^3} \left(1-(\hat{\mathbf{k}}\cdot\hat{\mathbf{q}})^2\right)\frac{d(|\mathbf{k}-\mathbf{q}|)}{(\mathbf{k}-\mathbf{q})^2\:\omega(q)}\\
 \label{G9.5}
\Delta I_\chi (k, \mu) & =  & I_\chi (k) - I_\chi (\mu)\hk ,\hk I_\chi(k) = \frac{N_c}{4}\int \frac{d^3q}{(2\pi)^3} \left(1-(\hat{\mathbf{k}}\cdot\hat{\mathbf{q}})^2\right) \frac{d(|\mathbf{k}-\mathbf{q}|)\:d(q)}{(\mathbf{k}-\mathbf{q})^2}\\
\label{G10}
 \Delta I^{(n)}_\omega (k, \mu) & = & I^{(n)}_\omega (k) - I^{(n)}_\omega (\mu)
 \hk , \hk n = 1, 2 \nonumber\\
 I^{(n)}_\omega(k) & = & \frac{N_c}{4}\int\frac{d^3q}{(2\pi)^3}\left(1+(\hat{\mathbf{k}}\cdot\hat{\mathbf{q}})^2\right)\nonumber\\
 & & \,\,\,\,\cdot\frac{d(|\mathbf{k}-\mathbf{q}|)^2f(|\mathbf{k}-\mathbf{q}|)}{(\mathbf{k}-\mathbf{q})^2}\cdot\frac{\left[\omega(q)-\chi(q)\right]^n-\left[\omega(k)-\chi(k)\right]^n}{\omega(q)}\hk .
 \eea

For the solution of the coupled Dyson-Schwinger equations (\ref{G7}) and (\ref{G8}),
we implement the horizon condition \cite{Zwa95}
\be
\label{G12}
d^{- 1} (\mu = 0) = 0 \hk .
\ee
Assuming power law behaviour for the Green functions in the infrared
\be
\label{G13}
\omega (k) = \frac{A}{k^\alpha} \hk , \hk d (k) = \frac{B}{k^\beta}\, ,
\ee
the infrared analysis of the ghost Dyson-Schwinger equation yields the sum rule
\be
\label{sum-rule}
2 \beta = \alpha + 1 \hk ,
\ee
due to the non-renormalization of the ghost-gluon vertex, a feature of
both Landau \cite{Tay71} and Coulomb gauge
\cite{FisZwa05,Schleifenbaum:2006bq}. The horizon condition (\ref{G12}) implies $\beta > 0$. For $\beta>1/2$ we obviously obtain an infrared
divergent gluon energy which is a manifestation of confinement. Furthermore,
$\beta > 1/2$ also implies an infrared divergent curvature $\chi (k)$.
Indeed, an infrared analysis \cite{Schleifenbaum:2006bq} of Eq.\ (\ref{G9.5}) shows that for $k\to 0$, $\chi(k) \propto k^{2\beta-1}$ which by the sum rule (\ref{sum-rule}) is the infrared behaviour of $\omega(k)$. This is consistent with the gap equation (\ref{eq-omega}). 
For an infrared divergent $\chi(k)$ the
gap equation (\ref{G8}) reduces in the infrared to 
\be
\label{G15}
\lim_{k\to 0}( \omega (k) - \chi (k) ) = c \hk ,
\ee
implying that $\omega (k)$ and $\chi (k)$ have the same infrared (divergent)
behaviour. Eq.\ (\ref{G15}) along with the ghost Dyson-Schwinger equation (\ref{G7}) in the
infrared limit can be solved analytically for the infrared
exponents. It was found earlier that, using angular approximation, one
finds only one solution $\beta=1$ \cite{Feuchter:2004mk}. However,
performing an infrared analysis without angular approximation, there are the two solutions \cite{Schleifenbaum:2006bq}
\be
\label{G16}
\beta_1 \approx 0.796 \, , \qquad \beta_2 = 1 \hk .
\ee
A numerical solution of the set of equations without the angular approximation was presented in ref.\ \cite{Feuchter:2004mk} corresponding to the value $\beta_1\approx 0.796$. In this paper we present an improved numerical method with which the infrared exponent $\beta_2=1$ is found as well.
\bi

After obtaining the self-consistent solutions for the form factors
numerically, we may apply them to the calculation of the static quark
potential on the one hand, and a non-perturbative running coupling on
the other. The static quark potential $V(r)$ can be computed by taking
the expectation value of the Coulomb term of the Yang-Mills Hamiltonian
(\ref{G1}) where $\rho$ is chosen to be the charge distribution of
two  heavy color charges, i.e.\
\begin{align}
\rho(x)=\delta^{(3)}(x-r/2)-\delta^{(3)}(x+r/2).
\end{align}
After Fourier transform, we arrive at
\begin{align}
\label{pot}
V(r)=\int \frac{d^3q}{(2\pi)^3}
\frac{d^2(q)f(q)}{q^2}\left(1-e^{i\mathbf{k}\cdot\mathbf{r}}\right)\, .
\end{align}
The solution with $\beta_2=1$, see Eq.\ (\ref{G16}), has the attractive
feature that it gives rise to a strictly linearly rising static quark
potential $V(r)=\sigma r$.

A non-perturbative running coupling may be extracted from the RG
invariant $V(r)$ given in Eq.\ (\ref{pot}) and would be divergent in
the infrared. At the same time, it is possible to define the running
coupling via the ghost-gluon vertex \cite{FisZwa05},
\begin{align}
\label{alpha}
\alpha(k)=\frac{16}{3}\frac{g_r^2}{4\pi}k^5\,G^2(k)D(k)=\frac{2}{3\pi}k\,d^2(k)\,\omega^{-1}(k)\, ,
\end{align}
and find that, due to the non-renormalization of the ghost-gluon
vertex, $\alpha(k)$ has an infrared fixed point. This can be
seen immediately when plugging in the sum rule (\ref{sum-rule}) into Eq.\
(\ref{alpha}). It was shown in ref.\ \cite{Schleifenbaum:2006bq} that
$\alpha(0)=\frac{16\pi}{3N_c}$ for $\beta=\beta_2$ by
infrared analysis of the Dyson-Schwinger equations. The same value
is actually found when resorting to the angular approximation
as done in \cite{Feuchter:2004mk}. There, it was shown that
$\frac{A}{B^2}=\frac{N_c}{6\pi^2}\frac{\beta+2}{2\beta(\beta+1)}$ which
yields the above mentioned infrared fixed point if $\beta=\beta_2$.

\bi

\no
\section{Numerical Solution of the Dyson-Schwinger Equations}
\bi

In order to solve the set of integral equations (\ref{G7}) and (\ref{G8})
numerically, we use a Chebyshev approximation on a logarithmic scale for the
form factors. A logarithmic momentum scale is introduced to have enough nodes at every order of magnitude. Furthermore, infrared- or ultraviolet divergent functions are expressed on a double-logarithmic scale.

Whenever we encounter a change of sign in one of the functions,
the logarithmic scale demands a separate treatment of positive and
negative values. In particular, the function $\chi(k)$
goes to $+\infty$ for $k\to 0$ and to $-\infty$ for $k\to\infty$,
with a change of sign at some intermediate point which we can fix by
the renormalization condition $\chi(\mu)=0$. This knowledge
then allows us to fit $\chi(k)$ logarithmically in the infrared, and
to fit $-\chi(k)$ also logarithmically in the ultraviolet. For some
range around $k=\mu$, we fit $\chi(k)$ linearly. The resulting set
of nodes is to be used also for the other form factors. This is done
to minimize the number of required Chebyshev evaluations since every
evaluation of a fit is subject to some non-zero error.

For asymptotic values of momentum, it is instructive for the numerics to make use of
the analytical results, cf.\ the methods used in ref.\ \cite{Bloch}. The set of nodes cover a finite momentum
region. In order to extrapolate this to the whole infinite momentum range, the general algebraic
forms obtained analytically serve as ans\"atze. However, contrary to what has been previously done in Dyson-Schwinger studies, the parameters of
these asymptotic forms are still determined numerically. E.g., in the infrared
it was shown analytically that power laws provide a solution. The numerics will have the asymptotic form of the power
laws (\ref{G13}) as an input but
determine the various exponents and coefficients as parameters
by nonlinear-least-squares fitting. Thus, this numerical method provides a check
on the analytical results (\ref{G16}) rather than imposing the
latter. 

Furthermore, since the set of integral equations comprises the
difference $\omega(k)-\chi(k)$ for small momenta $k$, we extract the
infrared power law $A/k^\alpha$ of $\omega(k)$ and $\chi(k)$ in a
single fit with the combined data which gives us $A$ and $\alpha$, and
extract the next-to-leading order constants $\omega_0$ and $\chi_0$,
again with least-squares fits. This gives us fits for $\omega(k)$
and $\chi(k)$ up to intermediate momentum values,
\begin{align}
\omega(k)= \frac{A}{k^\alpha}+\omega_0\, ,\qquad \chi(k)=
\frac{A}{k^\alpha}+\chi_0\, ,\quad k\rightarrow 0\, .
\end{align}
It is possible to gain more information on the intermediate momentum
regime by setting up an infrared expansion of the function (cf. Eq.\ (\ref{G15}))
\begin{align}
\label{defnu}
\nu(k):=\omega(k)-\chi(k)=c+c_1k^\gamma\, ,\quad k\rightarrow 0\, , \gamma>0\, .
\end{align}
and determining its parameters by least-squares fitting. However,
this works only if the form factors already are pretty close to the
(yet unknown) solution. Typically, at intermediate iteration stages
$\nu(k)$ is an infrared divergent quantity
(as opposed to the fully iterated solutions). It is then mandatory to
control that the fit of the function $\nu(k)$ converges to a sensible
value at each iteration step.

Ultraviolet extrapolation is done in a similar way. We know the
asymptotic behaviour from the UV analysis, see ref.\  \cite{Feuchter:2004mk}, and make corresponding
ans\"atze for the form factors in the UV,
\begin{align}
\omega(k\to\infty)&=k,& \chi(k\to\infty)&\propto k/\sqrt{\ln(k/\mu_\chi)}\, ,\\
d(k\to\infty)&\propto 1/\sqrt{\ln(k/\mu_d)},&f(k\to\infty)&\propto
1/\sqrt{\ln(k/\mu_f)}\, ,
\end{align}
extracting the coefficients as well as the different scale parameters $\mu_\chi$, $\mu_d$, $\mu_f$  using least-squares fitting.

Equipped with the discretized form factor ans\"atze, we can now
perform the numerical integration by quadrature followed by iteration
to determine the fit parameters. Here, we use the set of renormalized
and thus finite integral equations (\ref{G7}) and (\ref{G8}). Whereas the
angular integrals are performed with the customary Gauss-Legendre
formula, the radial integration involves additional non-linear
transformations. Having introduced finite infrared and ultraviolet
cutoffs $\varepsilon$ and $\Lambda$, resp., for the radial integrals
of the kind $\int_{0}^{\infty} dx\,f(x)$,\footnote{The error
  thus involved can be shown to be of the order
  ${\cal{O}}(\varepsilon)$ and ${\cal{O}}(1/\Lambda)$ to some positive
  power. Choosing $\varepsilon$ sufficiently small and $\Lambda$
  sufficiently large will render the error negligible.}
we transform the integration variable logarithmically to map the
integration interval to $[\ln\varepsilon, \ln\Lambda]$. Like in the
Chebyshev expansion, this logarithmic representation is chosen to have
enough nodes at every order of magnitude. The remaining integral
over this interval is evaluated numerically using a Gauss-Legendre
algorithm.

\begin{figure}
\begin{center}
\includegraphics{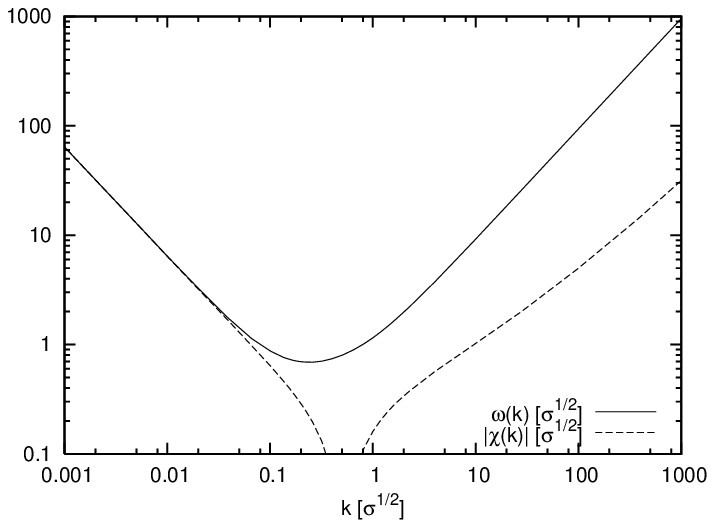}
\includegraphics{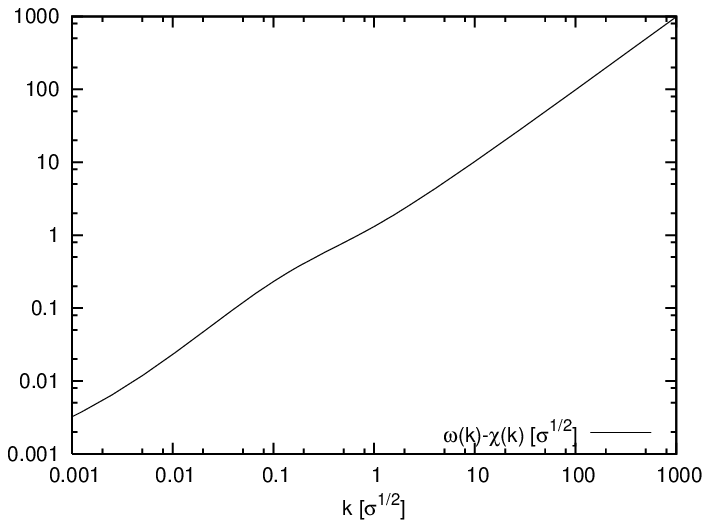}
\end{center}
\caption{Left panel: the gluon energy $\omega(k)$ and the modulus of the scalar curvature $\chi(k)$. Right panel: the difference $\omega(k)-\chi(k)$.\label{fig-omega-chi}}
\end{figure}

The numerical solution is then carried out iteratively,
thereby using some relaxation method (like $x_\mathrm{new} \leftarrow
rx_\mathrm{new}+(1-r)x_\mathrm{old}$) to compute the updated 
values for the form factors at the Chebyshev nodes. For a stable iteration we use
$r\approx 0.1$. These new values are then used to extract
the Chebyshev coefficients and the fit parameters for the infrared and ultraviolet
extrapolations. This gives us new form factors.

For the explicit calculation it is convenient to express the
Dyson-Schwinger equations in terms of dimensionless quantities by
rescaling all dimensionful quantities with appropriate powers of the
renormalization scale $\mu$. Indicating the rescaled quantities by bars,
one has
\begin{align}
\label{f71}
\bar{k} &= \frac{k}{\mu} \hk ,& \hk \bar{\omega} (\bar{k}) &=  \frac{\omega (k = \mu \bar{k})}{\mu} \hk ,& \hk \bar{d} (\bar{k}) &= d (k = \mu \bar{k}) \hk , \nonumber\\
& & \bar{\chi} (\bar{k}) &= \frac{\chi (k = \mu \bar{k})}{\mu} \hk ,& \hk \bar{c} &= \frac{c}{\mu} \hk .
\end{align}
Furthermore, we have to specify the renormalization constants, which
are $\bar{d}(1)$, $\bar{\omega}(1)$, $\bar{\chi}(1)$, $\bar{c}$. It
was already observed in ref.\ \cite{Feuchter:2004mk} that neither the
infrared exponents nor the ultraviolet behaviour of the various form factors
depend on the precise value of the renormalization constants except
for the value of $\bar{d} (1)$ which is, however, fixed  by the
horizon condition $d^{- 1} (0) = 0$. The remaining renormalization
constants are fixed as follows. Throughout our calculations we
choose $\bar{\chi} (1) = 0$. This choice is possible since from the
infrared and ultraviolet analysis of the Dyson-Schwinger equations
carried out in ref.\ \cite{Feuchter:2004mk} follows that $\chi (k)$
has a zero at some finite momentum value. Choosing $\bar{\chi}
(1) = 0$ has the advantage that the renormalization of the curvature
$\chi (k)$ just removes the ultraviolet divergence without introducing
any new finite parameter. In the same spirit we chose $c=0$ which implies
that $\omega(k)$ and $\chi(k)$ approach each other for $k\to 0$ (see Eq.\
(\ref{G15})). This choice is preferred by the study of the 't Hooft loop
\cite{ReToBePublished, Reinhardt:2006cf}. The remaining renormalization constant $\bar{\omega}(1)$
is considered as a parameter.

\begin{figure}
\begin{center}
\includegraphics{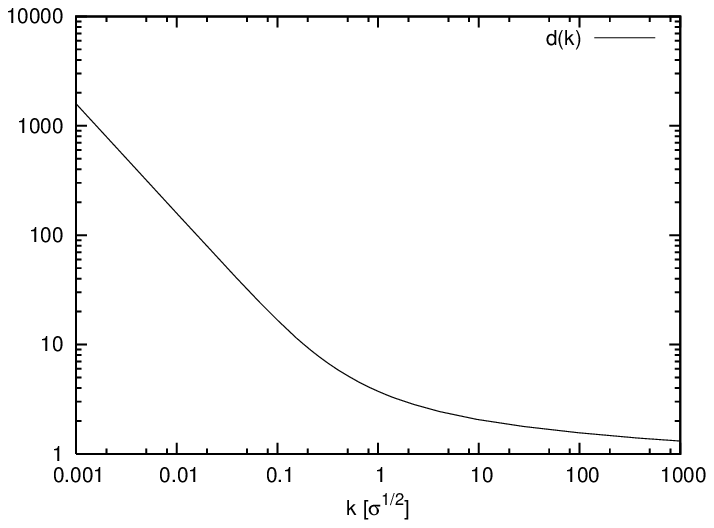}
\includegraphics{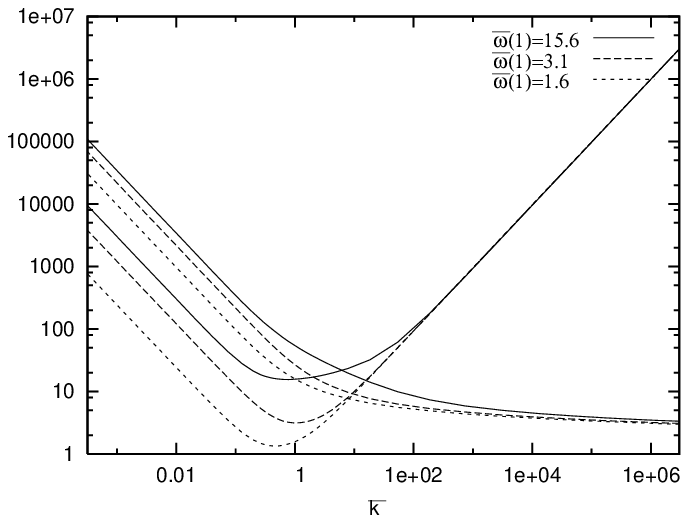}
\end{center}
\caption{Left panel: $d(k)$. Right panel: the functions $\omega(k)$ and $d(k)$
  for different values of $\omega(\mu)$.\label{fig-renorm}}
\end{figure}

Figures \ref{fig-omega-chi} and \ref{fig-renorm} show the
numerical solution of the coupled Dyson-Schwinger equations for
the choice $\bar{\omega} (1) = 1.6$. As seen from the left panel in
Fig.\ \ref{fig-omega-chi}, $\omega(k)$ and $\chi(k)$ indeed obey the same
infrared divergent behaviour, in agreement with the infrared analysis of the
Dyson-Schwinger equations \cite{Schleifenbaum:2006bq}. Furthermore from the right panel of Fig.\ \ref{fig-omega-chi} we also notice that in the infrared $\omega(k)$ and $\chi(k)$ coincide, in accord with our choice of the renormalization condition $c=0$.


Figure \ref{fig-renorm} shows the dependency of $\omega(k)$
and $d(k)$ upon the renormalization constant $\omega(\mu)$. We
find that this constant does not change the infrared exponents and the ultraviolet behaviour but does have impact on the prefactors $A$, $B$ of
the power laws for $\omega(k)$, $\chi(k)$ and $d(k)$.  Also, there
are some quantitative changes in the intermediate momentum range.

\begin{figure}
\begin{center}
\includegraphics{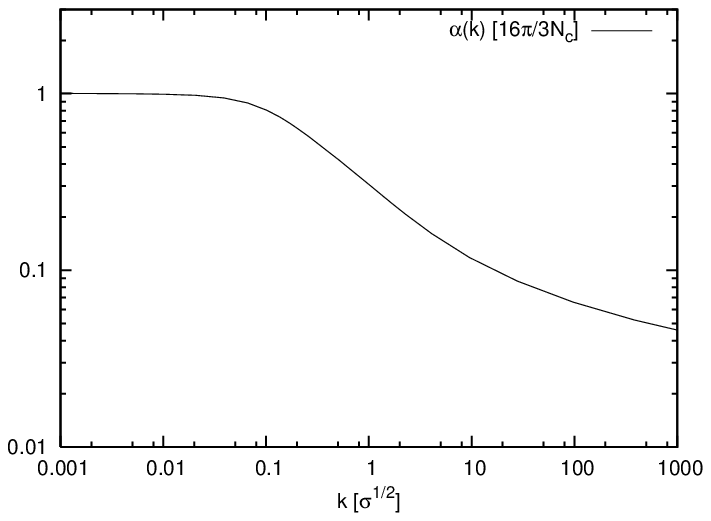}
\includegraphics{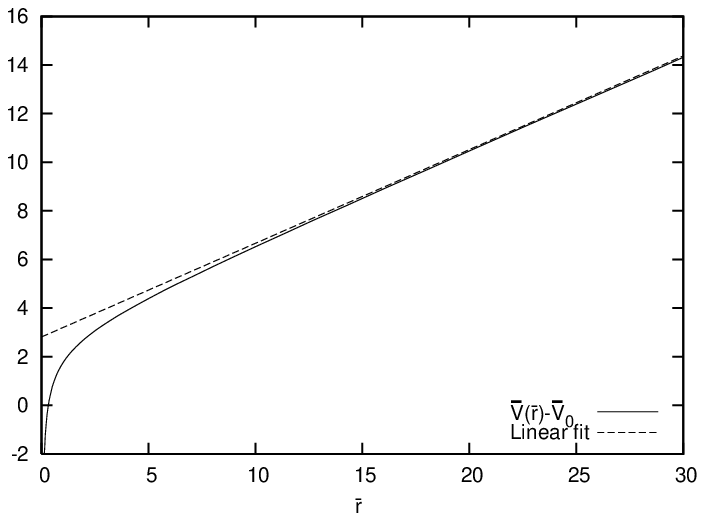}
\end{center}
\caption{(left) The running coupling $\alpha(k)$. (right) The Coulomb
potential $V(r)$.\label{fig-alpha-quark}}
\end{figure}

Figure \ref{fig-alpha-quark} (left panel) shows the running coupling,
as defined by Eq.\ (\ref{alpha}). It has an infrared fixed point which is in
excellent agreement with the value predicted by the infrared analysis
given in \cite{Schleifenbaum:2006bq}.
It is also worth mentioning that the running coupling
in Fig.\ \ref{fig-alpha-quark} is a monotonic function and has
no more than one inflection point and thus does not bring up peculiarities in the beta
function.

In the right panel of Figure \ref{fig-alpha-quark} we show the
Coulomb potential, as given by Eq.\ (\ref{pot}). We find an exactly
linearly rising behaviour (within an estimated error of less than
one percent), again in agreement with the infrared analysis \cite{Schleifenbaum:2006bq}. The linearly rising potential allows us to fit our scale from the string tension $\sigma$. Lattice calculations \cite{R8} show, however, that the Coulomb string tension is about a factor of $1.5\ldots 3$ larger than the string tension extracted from the Wilson loop, in agreement with the analytic result \cite{R9} that the Coulomb string tension is an upper bound to the Wilson loop string tension. 


\no
\section{Summary and Conclusions}
\bi

\no
We have numerically solved the Dyson-Schwinger equations following from
the minimization of the vacuum expectation value of the Yang-Mills
Hamiltonian in Coulomb gauge assuming the horizon condition. The
solution found produces a strictly linearly rising static quark potential which is in agreement with the analytic infrared analysis carried out
in ref.\ \cite{Schleifenbaum:2006bq}. We find only a rather weak dependence of our
solutions on the specific choice of the remaining renormalization constants,
in particular, the infrared exponents and the ultraviolet behaviour are independent
of the choices of the renormalization constants, once the horizon
condition is adopted. There are, however, changes in the intermediate
momentum range which, as will be shown in ref.\ \cite{ReToBePublished}, are tested by the 't Hooft loop.

\section*{Acknowledgments}
It is a pleasure to thank A.\ Szczepaniak for valuable discussions and
ongoing correspondence on the subject. We are grateful to the referee
for pointing out ref.\ \cite{Bloch} to us. This work
was supported by the Deutsche Forschungsgemeinschaft (DFG) under contract no.\
Re856/6-1,2 and by the Europ\"aisches Graduiertenkolleg Basel-Graz-T\"ubingen
(EUROGRAD).

\end{document}